\documentstyle[11pt]{article}

   \newcommand{\be}{\begin{equation}}
   \newcommand{\ee}{\end{equation}}
   \newcommand{\lb}{\label}

   \newcommand{\bv}{{\bf v}}
   \newcommand{\br}{{\bf r}}
   \newcommand{\bk}{{\bf k}}
   \newcommand{\bl}{{\bf l}}
   \newcommand{\bh}{{\bf s}}
   \newcommand{\bj}{{\bf j}}

   \newcommand{\bs}{{\bf s}}

   \newcommand{\bx}{{\bf x}}

   \newcommand{\vl}{\overline{{\bf v}}}
   \newcommand{\vs}{{\bf v}^{\prime}}

   \newcommand{\pl}{\overline{p}}
   \newcommand{\el}{\overline{e}}
   \newcommand{\btau}{{\mbox{\boldmath $\tau$}}}
   \newcommand{\bdot}{{\mbox{\boldmath $\cdot$}}}
   
   \newcommand{\grad}{{\mbox{\boldmath $\nabla$}}}
   \newcommand{\bsigma}{{\mbox{\boldmath $\sigma$}}}

\begin{document}
   \title{ The `Multifractal Model' of Turbulence and {\em A Priori} Estimates in
   Large-Eddy Simulation, I.
	   Subgrid Flux and Locality of Energy Transfer}
   \author{Gregory L. Eyink\\{\em Department of Mathematics, University of
   Arizona}\\{\em Tucson, AZ 85721}}
   \date{ }
   \maketitle
   \begin{abstract}
   We establish and discuss {\em a priori} estimates on subgrid stress and subgrid
   flux for filtering schemes used in the turbulence modelling method of 
   Large-Eddy Simulation (LES). Our estimates are derived as rigorous consequences 
   of the exact subgrid stress formulae from Navier-Stokes equations under 
   realistic conditions for inertial-range velocity fields, those conjectured in 
   the Parisi-Frisch ``multifractal model.'' The estimates are shown to be an 
   expression of ``local energy cascade,'' i.e. the dominance of local wavevector
   triads in the energy transfer. We prove that for nearly any reasonable filter
   function the LES method defines an energy flux in which local triads dominate 
   in individual realizations, due to cancellation of distant triadic
   contributions by detailed conservation. A somewhat similar observation of
   Leslie and Quarini on graded filters in the EDQNM closure is shown to be 
   unrelated to the cancellation we establish in Navier-Stokes solutions.
   The sharp Fourier cutoff filter is one example which does not satisfy the
   modest conditions of our proof and, in fact, we show that with that filter 
   the energy transfer in individual realizations at arbitrarily high Reynolds
   number will be dominated by nonlocal, convective sweeping.


   \end{abstract}

   \newpage

   \section{Introduction}

   A fundamental hypothesis in the Kolmogorov phenomenology of developed
   turbulence is the ``local
   energy cascade'' \cite{1a,1b,1c}. In the first of his 1941 papers, Kolmogorov
   expressed this in spatial
   terms as a process of energy transport to small scales by successive transfer
   from ``pulsations''
   of one size to daughter ``pulsations'' of size about half as large. The
   parallel works of Heisenberg \cite{2},
   von Weizs\"{a}cker \cite{3} and Onsager \cite{4a,4b} stated this in spectral
   terms as a process in which
   energy is transported through Fourier space along chains of wavevector triads
   in which typically
   two wavevectors of the same size transfer energy to a third wavevector about
   twice as big. Recently,
   the dominance of these local wavevector triads to turbulent energy transfer has
   been the subject
   of much debate. The papers \cite{5}-\cite{17} in chronological order are a
   representative sample containing many further
   references. The cited studies are based mostly upon direct numerical simulation
   of the Navier-Stokes equations
   or upon the solution of analytical closures, such as the EDQNM transfer
   equation. The first method
   is severely constrained by the low Reynolds numbers attainable with current
   computer resources.
   The numerical solution of analytical closures, on the other hand, is feasible
   at high Reynolds number---
   at least in the case of homogeneous and isotropic turbulence---but is subject
   to doubt concerning the
   uncontrolled closure hypotheses invoked.

   However, we have in fact demonstrated in our past work \cite{18} locality of
   instantaneous energy
   transfer for individual turbulent solutions of the Navier-Stokes equations
   based upon a direct,
   rigorous analysis of the inertial terms and thus valid for Reynolds numbers
   arbitrarily large
   without any assumption of statistical homogeneity or isotropy. Our original
   study involved
   mathematical considerations on so-called ``weak solutions'' which might make it
   inaccessible to fluid-dynamicists
   without a good grounding in modern analysis. Therefore, we shall present here
   the basic arguments of
   our earlier work with an emphasis on their physical interpretation in order to
   place them in the
   context of the existing turbulence literature. In addition, we shall discuss
   stronger results
   on simultaneous space-scale locality of transfer which we subsequently obtained
   \cite{19} and new material
   exposing the various relations between both of our two works and the previous
   literature.

   We have found that the filtering technique commonly used in the Large-Eddy
   Simulation or LES
   modelling method (see \cite{20,21}) provides the most theoretically
   well-founded
   definition of the intuitive idea of ``scales of motion'' in turbulent flow. In
   this method a
   filtering function $G_\ell(\br)=\ell^{-d}G_\ell(\br/\ell)$ is convoluted with
   the velocity field, as
   \be \vl_\ell(\br)\equiv \int d^d\br'\,\,G_\ell(\br-\br')\bv(\br'), \lb{1} \ee
   to define the ``large-scale velocity'' or ``resolved field'' $\vl_\ell.$ The
   ``small-scale velocity'' or
   ``subgrid field'' is then simply defined as the complementary component:
   \be \vs_\ell(\br)\equiv \bv(\br)-\vl_\ell(\br). \lb{2} \ee
   When the filtering operation is applied to the Navier-Stokes system an equation
   is obtained for $\vl_\ell$:
   \be \partial_t \vl_\ell+\nabla\cdot(\vl_\ell\vl_\ell+\btau_\ell)=-\nabla
   \pl_\ell+\nu_0 \bigtriangleup \vl_\ell, \lb{3} \ee
   in which $\pl_\ell$ is the filtered pressure field (required to maintain
   $\nabla\cdot\vl_\ell=0$) and
   \be \btau_\ell=\overline{ (\bv\bv)}_\ell-\vl_\ell\vl_\ell, \lb{4} \ee
   is a tensor representing the ``subscale stress'' of the eliminated turbulent
   eddies.

   We shall use the same filtering technique as in conventional LES but in order
   to derive rigorous mathematical
   estimates for the exact subgrid stress $\btau_\ell$ which follows from
   Navier-Stokes. We show
   under very mild assumptions on the filter function $G$ that the energy flux
   which appears in the LES method is dominated
   by local wavevector triads, in agreement with the conventional wisdom. However,
   a few commonly used filters---
   including the sharp cutoff filter in Fourier space--- will be shown to violate
   these assumptions and to lead to
   formulae for subscale stress which are not appropriate to model as an effect of
   the small-scales alone. The LES technique
   with a ``good''' filter will in fact be shown to simply and naturally resolve
   certain subtleties of the locality issue,
   discussed in \cite{5}-\cite{17}, which arise purely as an artefact of the
   conventional use of Fourier series.

   Our results have also some further implications. First, the derived estimates
   place an {\em a priori} constraint on
   subgrid models and filters and may be used to evaluate the various choices.
   This is the subject of a companion paper
   \cite{I2}, hereafter referred to as II. Second, the estimates imply
   restrictions on the space-regularity of turbulent
   velocity fields in the inertial range of scales in order to be consistent with
   the basic phenomenology of constant mean
   energy flux. Roughly speaking, the velocity must appear on those scales as a
   continuous, nowhere-differentiable field of
   the type represented by the famous Weierstrass function or a typical path of
   Brownian motion. This was noted already by
   Onsager in 1949 \cite{4b}.

   In fact, all of the estimates we derive are based upon the regularity of
   inertial-range velocity fields postulated in the
   Parisi-Frisch ``multifractal model'' \cite{22}, who generalized Onsager's
   observation. See \cite{23},\cite{24} for recent
   accounts. Let us just recall here that those authors proposed that in the limit
   of zero viscosity the velocity field in
   turbulence should be {\em H\"{o}lder continuous} at each point, that is, for
   every point $\br$ in the flow domain an
   inequality
   \be |\bv(\br+\bl)-\bv(\br)|\leq C|\bl|^h \lb{5} \ee
   should be satisied for some real $h$ and all $|\bl|\leq 1.$ (This definition
   must be modified for negative $h.$) If so,
   then to each point may be assigned the largest value of $h$ for which
   Eq.(\ref{5}) holds and it was furthermore postulated
   that the points where the value $<h$ occurs should be a fractal set $S(h)$ with
   dimension $D(h).$ This is the origin
   of the word ``multifractal.'' Notice that the original 1941 Kolmogorov theory
   corresponds in this language to $D(1/3)=3$
   and $D(h)=-\infty$ otherwise, that is, $h=1/3$ everywhere. \footnote{The
   original motivation for this ``multifractal
   hypothesis''was the experimental observation \cite{25} that longitudinal
   velocity-difference moments (structure functions)
   scale as powers of the separation length $\ell=|\bl|,$
   \begin{eqnarray}
   S_p(\ell) & \equiv & \langle \left[\hat{\bl}\cdot\left(\bv (\br +\bl
   )-\bv(\br)\right)\right]^p\rangle \cr
	  \, & \sim   & v_0^p\left({{\ell}\over{L}}\right)^{\zeta_p}, \lb{6}
   \end{eqnarray}
   over inertial-range scales $\eta\ll \ell\ll L$ and that $\zeta_p\neq p/3,$ the
   prediction of K41 theory.
   Parisi and Frisch explained this latter fact heuristically by noting that, with
   their hypotheses,
   \be S_p(\ell)\sim v_0^p \int
   d\rho(h)\,\,\left({{\ell}\over{L}}\right)^{ph}
	     \cdot\left({{\ell}\over{L}}\right)^{3-D(h)}, \lb{7} \ee
   where the second factor measures the fraction of boxes of size $\ell$ in the
   domain of size $L$ in which the exponent $h$ occurs.
   Applying steepest descent to evaluate this integral for $\ell\ll L$ they
   obtained Eq.(\ref{6}) with
   \be \zeta_p=\inf_h[ph+(3-D(h))]. \lb{8} \ee}
   Although the model must be considered to be conjectural, we shall see that it
   has some mathematical
   support from the Navier-Stokes equations. In fact, the ``multifractal''
   velocities are the most regular velocity-fields
   possible which are still compatible with the basic physics of constant mean
   energy flux through inertial-range scales.
   In particular, it is not possible for the velocity to be everywhere
   space-differentiable on inertial-range length-scales
   (which would imply that Eq.(\ref{5}) hold at all points $\br$ with $h\geq 1.$)
   Just to be clear, we emphasize that
   the conclusion is that Eq.(\ref{5}) cannot hold for Navier-Stokes solutions
   with $h>1/3$ at all $\br$ and
   inertial-range $\ell.$ The argument for this is simple, but somewhat off the
   main line of the paper, so that
   we include it as an Appendix.

   The contents of our paper are as follows: In the next Section II we shall
   derive {\em a priori} estimates on
   subgrid stress and flux as defined in the LES method, which are {\em exact}
   consequences of Navier-Stokes assuming
   the ``multifractal'' regularity of velocities. In Section III we explain how
   the derived
   estimates express the ``locality'' of the instantaneous energy-transfer. It is
   shown by explicit analysis that for a
   ``good'' choice of filter the LES flux gets a negligible contribution from
   distant wavenumber triads and we thereby
   resolve some of the controversies regarding ``locality'' of transfer. We
   contrast our result with an earlier
   observation of Leslie and Quarini \cite{LQ}. We also show that there are
   ``bad''
   choices of filter for which the flux is dominated by distant triadic
   interactions of convective type.
   Finally, in Section IV  we make some closing remarks and conclusions on the
   ``multifractal model,'' the locality
   of transfer, and the filtering approach.

   \newpage

   \section{ {\em A Priori} Estimates and the ``Multifractal Model''}

   We have already introduced the concept of ``subscale stress'' $\btau_\ell$ in
   the LES
   filtering technique. It is also natural to write down a {\em local energy
   balance}
   for the resolved kinetic energy
   \be \el_\ell(\br,t)\equiv {{1}\over{2}}\vl_\ell^2(\br,t), \lb{9} \ee
   as
   \be \partial_t
   \el_\ell(\br,t)+\nabla\cdot\bj_\ell(\br,t)
		  =-\Pi_\ell(\br,t)-\varepsilon_\ell(\br,t). \lb{10} \ee
   For example, see Section 6.1 of \cite{21}.  Each of the terms in the
   microscopic balance relation Eq.(\ref{10})
   has a precise physical interpretation. The term $\varepsilon_\ell(\br,t)\equiv
   \nu_0 (\nabla\vl_\ell)^2$ represents
   the local dissipation of energy in the length-scales $>\ell$ through the action
   of molecular viscosity. The term
   $\bj_\ell(\br,t)\equiv
   (\pl_\ell+\el_\ell)\vl_\ell+\btau_\ell\cdot\vl_\ell-\nu_0\nabla \el_\ell$ is a
   spatial current
   of energy in the scales $>\ell.$  Finally, the term of main interest for us
   \be \Pi_\ell(\br)\equiv -(\nabla\vl_\ell):\btau_\ell \lb{12} \ee
   represents a {\em local energy flux} from the large-scales to the small-scales.
   It gives the ``effective dissipation''
   of energy in the large-scales $>\ell$ due to the action of the eddy-stress of
   the small-scales $<\ell$ on the gradients
   of the large-scale motion.

   In this section we shall derive some simple {\em a priori} estimates for the
   quantities $\btau_\ell(\br)$ and
   $\Pi_\ell(\br)$ assuming the H\"{o}lder regularity Eq.(\ref{5}) at point $\br,$
   a condition usually denoted
   $\bv\in C^h(\br).$ Only afterward will we consider the consequences of the
   estimates and the justification
   for the assumed regularity. A key identity which was noted, in a different
   language, in a paper of Constantin et al.
   \cite{27}, is
   \be \btau_\ell(\br)=\langle\Delta\bv(\br)\Delta\bv(\br)\rangle_\ell
			 -\vs_\ell(\br)\vs_\ell(\br). \lb{14} \ee
   where
   \be \Delta_\bh\bv(\br)\equiv \bv(\br)-\bv(\br-\bh) \lb{15} \ee
   is the ``backward difference'' and
   \be \langle f\rangle_\ell\equiv \int d\bh \,\,G_\ell(\bh) f(\bh), \lb{16} \ee
   denotes an ``average'' over the separation distance with respect to the filter
   function.
   The proof is elementary. Writing the righthand side of Eq.(\ref{14}) as
   \be \int d\bh \,\,G_\ell(\bh)
   [\bv(\br)-\bv(\br-\bh)][\bv(\br)-\bv(\br-\bh)]
	-[\bv(\br)-\vl_\ell(\br)][\bv(\br)-\vl_\ell(\br)], \lb{17a} \ee
   using the normalization condition
   \be \int d^d\br\,\,G_\ell(\br)=1 \lb{18} \ee
   and some shifting of integration variables and cancelling terms gives
   $\overline{(\bv\bv)}_\ell-\vl_\ell\vl_\ell,$
   which is the definition of the subscale stress.

   A more symmetric and suggestive form of the same identity is
   \be \btau_\ell(\br)=\langle\Delta\bv(\br)\Delta\bv(\br)\rangle_\ell
   -\langle\Delta\bv(\br)\rangle_\ell\langle\Delta\bv(\br)\rangle_\ell. \lb{17}
   \ee
   The proof is instructive. By definition
   \be \vl_\ell(\br)=\int d^d\br'\,\, G_\ell(\br-\br')\bv(\br'). \lb{19} \ee
   Therefore, using $\bv_\ell'=\bv-\vl_\ell$ and the normalization condition
   Eq.(\ref{18}) for $G_\ell,$
   \begin{eqnarray}
   \bv_\ell'(\br)&=& \int d^d\br'\,\,G_\ell(\br-\br')[\bv(\br)-\bv(\br')] \cr
	       \,&=& \int d^d\bh
   \,\,G_\ell(\bh)\Delta_\bh\bv(\br)\equiv\langle\Delta\bv(\br)\rangle_\ell.
   \lb{20}
   \end{eqnarray}
   Therefore, the subgrid velocity $\vs_\ell$ is represented by an ``average''
   over velocity-differences with
   separation $<\ell.$ Substituting into the previous formula Eq.(\ref{14}) for
   $\btau_\ell$ gives Eq.(\ref{17}).
   In this expression for the subscale stress the velocity again appears only
   through its difference $\Delta\bv$
   over length-scales $<\ell.$ If the filter function $G$ is positive as well as
   normalized, then Eq.(\ref{16})
   defines a true average and, from Eq. (\ref{17}), $\btau_\ell$ is a positive,
   symmetric tensor. An equivalent
   identity was discovered and used to establish positivity of the stress in a
   recent work of Vreman et al. \cite{VGK},
   which appeared after the first submission of this paper.

   We can now straightforwardly derive our {\em a priori} estimates using the
   condition $\bv\in C^h(\br),$ i.e.
   $|\Delta_\bh\bv(\br)|\leq ({\rm const.})|\bh|^h. $ For example, it follows
   directly from Eq.(\ref{20}) that
   \begin{eqnarray}
   |\bv'_\ell(\br)| &\leq& ({\rm const.})\int d^d\bh \,\,G_\ell(\bh)|\bh|^h \cr
		    & = & O\left(\ell^h\right). \lb{22}
   \end{eqnarray}
   if the filter function satisfies the modest condition $\int
   d\bx\,\,|G(\bx)|\cdot|\bx|^h<\infty.$ Likewise,
   \be \btau_\ell(\br)=O\left(\ell^{2h}\right) \lb{24} \ee
   by using Eq.(\ref{17}) if $\int d\bx\,\,|G(\bx)|\cdot|\bx|^{2h}<\infty.$ This
   is our main estimate on the subscale stress.
   To obtain an estimate on the subgrid flux $\Pi_\ell$ we also require a formula
   for $\nabla\vl_\ell$ expressing it in terms
   of velocity-differences over length-scales $<\ell.$ Using $\int\,\nabla G=0$ we
   find
   \begin{eqnarray}
   \nabla\vl_\ell(\br) &=& \int d^d\br'\,\,(\nabla
   G_\ell)(\br-\br')\left[\bv(\br')-\bv(\br)\right] \cr
		       &=& -\int d^d\bh \,\,(\nabla
   G_\ell)(\bh)\Delta_\bh\bv(\br). \lb{26}
   \end{eqnarray}
   Making the same estimations as before, we then find that
   \be \nabla\vl_\ell(\br)=O\left(\ell^{h-1}\right), \lb{27} \ee
   if $\int d\bx\,\,|\nabla G(\bx)|\cdot|\bx|^h<\infty.$ From this result and the
   formula $\Pi_\ell=
   -\nabla\vl_\ell:\btau_\ell$ it follows that
   \be \Pi_\ell(\br)=O\left(\ell^{3h-1}\right). \lb{29} \ee
   This is our main estimate for the flux.

   If we gather together all of the conditions on the filter function $G,$ we see
   that it is sufficient
   for the previous estimates to apply that it obey the moment condition
   \be \int d\bx\,\,\left[|G(\bx)|+|\nabla G(\bx)|\right]\cdot|\bx|^{2}<\infty,
   \lb{30} \ee
   when the H\"{o}lder indices lie in the range $0<h<1.$ The condition is quite
   modest. Of filters
   commonly used in practical LES, the {\em Gaussian filter} is a ``good'' example
   which easily satisfies
   these constraints. On the other hand, another common filter, the {\em sharp
   cutoff filter},
   violates condition Eq.(\ref{30}), because of slow spatial decay in space $\sim
   |\bx|^{-d}$ for large $\bx.$
   In fact, we will see by an example from \cite{18} that the previous bounds for
   $\btau_\ell$
   and $\Pi_\ell$ may actually be violated for this filter. Smoothly cutting off
   the tails of the filter in
   physical space would cure this problem in principle, but there are definite
   signs of the bad behavior in
   practical use of the sharp-cutoff filter in current LES simulations. This is
   discussed at length in II.

   So far we have employed the hypothesis of local regularity made in the
   Parisi-Frisch ``multifractal model,''
   without considering its validity. Since the assumption of local H\"older
   regularity cannot presently
   be derived {\em a priori} as a theorem for solutions of the 3D Euler equations,
   it does require some justification.
   Mathematical results---presented in \cite{38}---in conjunction with basic facts
   of turbulence phenomenology
   strongly suggest the ``multifractal model'' as the correct candidate to
   describe the space variation of inertial-range
   velocity fields. We must emphasize, however, that the key elements of the
   ``multifractal model'' used in our previous
   analysis {\em do not depend upon the questionable experimental results on
   anomalous scaling} \cite{25}. On the contrary,
   much weaker conditions guarantee the validity of the local H\"{o}lder
   regularity we employed. A particular condition which
   suffices is that the inertial-range energy spectrum be power-law $E(k)\sim
   k^{-n}$ with any spectral exponent whatsoever.
   In fact, we have proved elsewhere \cite{38} that {\em any homogeneous random
   field with such a power-law
   spectrum has realizations with local H\"{o}lder regularity with probability
   one.}  We just emphasize here that
   ``intermittency'' is irrelevant for this issue and that the spectral exponent
   could be the Kolmogorov one, $n=5/3,$
   or anything else. \footnote{If the energy spectral exponent is exactly
   $n={{5}\over{3}},$ then
   ``negative H\"{o}lder singularities'' may occur but can occupy a space set of
   Hausdorff dimension
   equal to {\em at most} $2{{1}\over{3}}$ \cite{38}. If $n>{{5}\over{3}},$ then
   the Hausdorff dimension of the
   ``negative exponent'' set is even smaller.} Furthermore, local H\"{o}lder
   regularity is supported by the estimate
   Eq.(\ref{29}), which implies $\Pi_\ell\rightarrow 0$ as $\ell\rightarrow 0$ if
   $h>{{1}\over{3}}.$ It is impossible to
   account for mean energy flux through the inertial range if that condition holds
   (e.g. if the velocities are spatially
   smooth) at all points of the domain on inertial-range scales. Therefore, {\em
   the ``multifractal model'' hypothesis
   of local H\"{o}lder regularity is vital to account for turbulent energy balance
   in the inertial-range.}

   \section{ Locality of Energy Transfer}

   \noindent {\em 3.1 The Estimates and Local Transfer in Wavenumber}

   The {\em a priori} estimates on subscale stress and energy flux in LES were
   derived in the previous section
   just as mathematical upper bounds. However, they have a very intuitive physical
   interpretation
   in terms of {\em locality of energy transfer}, which we discuss in this
   section. An essential point
   in the LES method is that it defines, via Eq.(\ref{12}), an energy flux which
   is proportional only to the {\em gradient}
   of the large-scale velocity field. The physics of this expression seems
   transparent: it describes the energy transfer
   due to the stretching action of the large-scale strain
   $\overline{\bsigma}_\ell\equiv
   {{1}\over{2}}\left(\nabla\vl_\ell+\nabla\vl_\ell^\top\right)$
   upon the stress field $\btau_\ell(\br)$ due to the random distribution of
   vorticity elements in the small-scales $<\ell.$
   In fact, it is often convenient to regard the fine-grained turbulence as a
   ``tangle'' of vortex filaments with a viscoelastic
   response to imposed large-scale strain \cite{41}. All of the {\em explicit}
   contributions of convective processes to the rate
   of change of the resolved energy $\el_\ell(\br,t)$ at the point $\br$ in the
   LES method are incorporated into the term
   $(\el_\ell+\pl_\ell)\vl_\ell$ of the current $\bj_\ell$ and can be accounted as
   contributions to transport of energy in
   space rather than ``downward'' to smaller scales. Since it is the large-scale
   convective processes which can produce apparent
   violations of locality in energy transfer in the wavenumber description, it is
   rather natural to find that $\Pi_\ell(\br)$
   obeys a ``local estimate'' $\sim \ell^{3h-1}$ which arises from
   $\nabla\vl_\ell\sim \ell^{h-1}$ and
   $\btau_\ell\sim \vs_\ell\vs_\ell\sim \ell^{2h}.$

   However, we shall show that this simple picture actually depends upon the
   technical-looking condition Eq.(\ref{30})
   we imposed upon the filter function $G$ in the previous section. Contrary to
   naive expectations, the LES flux can retain
   {\em implicit} contributions from purely convective processes if that (mild)
   condition is violated. We shall show this by a
   concrete example. In such cases it is inappropriate to estimate $\btau_\ell\sim
   \vs_\ell\vs_\ell$ because the ``subscale
   stress'' is actually still dependent upon the large-scale motion! In the course
   of our analysis we will make contact with
   the traditional discussion of the locality issue in Fourier space and resolve
   the subtle points which arise in that
   description.

   In the Fourier representation, the conventional definition of ``(subscale)
   energy flux'' $\Pi(k,t)$
   goes back at least to the work of Heisenberg \cite{2} and von Weizs\"{a}cker
   \cite{3}, who defined it as
   the total output power from the Fourier modes in the sphere of radius $k$ in
   wavenumber space:
   \be \Pi(k,t)=\left. -{{dE^{<k}(t)}\over{dt}}\right|_{{\rm Euler}}.  \lb{65} \ee
   Here $E^{<k}(t)$ is the instantaneous energy in modes of wavenumber $<k.$ If
   the {\em sharp cutoff filter} is used in
   LES in a rotationally symmetric form, as
   \be \widehat{G}(\bk)=\left\{ \begin{array}{ll}
		       1 & \mbox{ if $|k|<2\pi$} \cr
		       0 & \mbox{ otherwise.}
		       \end{array}
	       \right., \lb{68} \ee
   then the global flux thereby defined
   \be \Pi_\ell=\int_\Omega d\br\,\,\Pi_\ell(\br) \lb{69} \ee
   coincides with the traditional flux, through the relation
   $\Pi_\ell=\Pi(2\pi/\ell).$

   For a ``good'' filter obeying the condition Eq.(\ref{30}) it follows from the
   (stronger) local estimate of
   the previous section that $\Pi_\ell=O\left(\ell^{3h-1}\right)$ if the minimum
   H\"{o}lder exponent of the velocity is $h.$
   However, it was found in \cite{18} that the global flux for the sharp Fourier
   filter may be much larger, and, in general,
   only the estimate
   \be \Pi(k)=O\left(k^{1-2h}\right), \lb{F1} \ee
   may be true. In Example 1 of \cite{18} we constructed in 3D a velocity field
   $\bv\in C^h(\Omega)$, $0<h<1,$ for which
   $\Pi(k)=({\rm const.})k^{1-2h}$ at a set of $k$'s of arbitrarily
   large magnitude (namely, $k=2^\Lambda k_0$ with integer $\Lambda>0.$) It is
   important
   here just to recall the salient features of that example. We have already seen
   that the estimate $O(2^{\Lambda(1-3h)})$
   is true if the local interactions dominate in the energy flux. Therefore, the
   example constructed was of the opposite
   sort, for which highly nonlocal triads are dominant. The idea of the
   construction was very simple: we
   used a Fourier series in which every octave band of wavevectors contained
   one conjugate pair of modes at the very bottom and one pair just below the top,
   in just such a way that the top pair of each band and the bottom pair of the
   next highest band differ by a fixed pair of wavevectors in the {\em lowest}
   band.
   This pair of low-wavenumber modes induce an instantaneous transfer of energy
   between the
   high-wavenumber bands from the pairs at the top of one band into those at the
   bottom of the
   next higher band, without losing any energy themselves. This type of
   ``catalytic'' effect of the low-wavenumber
   modes for transfer between high-wavenumbers was, to our knowledge, first
   discussed in detail by Brasseur and Corssin
   in \cite{5}. The same effect was observed in the numerical study of Domaradzki
   and Rogallo \cite{7}, who
   described it as ``local transfer by nonlocal triads'' (although their
   simulation was at a rather low
   Reynolds number and did not make clear whether such interactions might persist
   as an inertial-range effect). The triads
   involved correspond to an ``R-interaction'' in the classification of Waleffe
   \cite{11}.

   A remarkable feature of the example is that the flux is proportional to the
   {\em amplitude} $U$ of the
   low-wavenumber modes rather than to their gradient. It therefore corresponds to
   an interaction of
   strictly convective type. This interpretation of the ``R-interactions'' was
   also argued by
   Waleffe in \cite{11}, whereas Domaradzki \cite{14} has put forward a very
   different picture
   ``...where the role of the large scales is to produce intermittent regions  of
   relatively strong,
   internal shears characterized by smaller length scales, which serve as regions
   of efficient small-scale transfer.''
   The explicit verification of the convective nature of these processes renders
   the latter
   explanation untenable for the inertial-range at high Reynolds number. In fact,
   the basic physics
   of these processes was already discussed by Kraichnan in 1966 in the context of
   the energy transfer
   equation of ALHDIA \cite{42}. He noted that: ``Convection of high-wavenumber
   structures by strongly
   excited low-wavenumber velocity components implies a rapid change of phase of
   the high-wavenumber
   Fourier amplitudes; that is to say, a rapid exchange of energy between sine and
   cosine components
   of the high wavenumbers. This exchange is represented in (2.6) [his transfer
   equation] by the large,
   cancelling input and output contributions. The net contribution $T_{<q}(k,t)$
   represents the effect
   of straining alone.'' In other words, the convection of structures at a high
   wavenumber $\sim k$
   by the largest, most energetic eddies will result in a rapid transfer of energy
   by a small distance in
   wavenumber back and forth across the sharp spectral boundary at $k.$ In a
   time-average of the flux
   $\langle \Pi(k)\rangle$ such effects may be expected to cancel, not only within
   the DIA closure,
   but in actual fact. However, for an individual ensemble realization at a given
   instant in time such
   effects will be present in $\Pi(k),$ and, indeed, will entirely dominate in
   $\Pi(k)$ at large $k,$
   swamping the comparatively small contribution of local triads.

   Although the nonlocal, convective effects may be expected to cancel in a time-
   or ensemble-average of
   $\Pi(k),$ this would hardly suffice to justify Kolmogorov's strong statement of
   statistical independence
   of the small-scale modes. In fact, the previous observations have raised some
   doubts, voiced in \cite{7,8,12},
   whether the small scales will really achieve such independence, e.g. of the
   external forcing or of anisotropy
   of the large scales. However, it seems clear from our discussion above that the
   nonlocal contributions
   to $\Pi(k)$ are an artefact of the sharp spectral boundary, and do not
   represent an actual physical mechanism
   for transfer of excitation from large scales to small ones. We have already
   seen that the LES flux, with
   a ``good'' filter choice, gets no net contribution from such terms even at a
   single instant in a fixed flow
   realization (when the H\"{o}lder regularity is satisfied). Instead, this flux
   properly represents energy transfer
   as due to the large-scale strain. It is worthwhile to study in more detail the
   wavevector contributions to
   the LES flux in order to understand the origin of cancellations in the nonlocal
   triads.

   The relation of the LES flux to the conventional spectral flux $\Pi(k)$ is
   obtained very easily when the
   filter function $G$ is spherically symmetric, as we hereafter assume. If the
   total resolved kinetic
   energy is represented as
   \begin{eqnarray}
   \overline{E}_\ell(t) & \equiv & \int d\br\,\,{{1}\over{2}}\vl_\ell^2(\br,t) \cr
		      \,& =      & \int
   {{d\bk}\over{(2\pi)^d}}\,\,{{1}\over{2}}|\hat{\bv}(\bk,t)|^2|
					    \widehat{G}(k\ell)|^2, \lb{86}
   \end{eqnarray}
   then using $\Pi_\ell= -\left.d\overline{E}_\ell/dt\right|_{{\rm Euler}},$ it
   follows that
   \begin{eqnarray}
   \Pi_\ell & = & \int_0^\infty dk\,\,{{\partial\Pi(k)}\over{\partial
   k}}|\widehat{G}(k\ell)|^2 \cr
	  \,& = & \int_0^\infty dk\,\, \Pi(k)D_\ell(k), \lb{87}
   \end{eqnarray}
   where
   \be D_\ell(k)\equiv -{{\partial}\over{\partial k}}|\widehat{G}(k\ell)|^2.
   \lb{88} \ee
   Observe that
   \begin{eqnarray}
   \int_0^\infty dk\,\, D_\ell(k) & = & |\widehat{G}(0)|^2-|\widehat{G}(\infty)|^2
   \cr
				\,& = & 1-0=1, \lb{89}
   \end{eqnarray}
   and from its definition Eq.(\ref{88}) that
   \be D_\ell(k)\geq 0, \lb{90} \ee
   if $|\widehat{G}(k)|^2$ decays monotonically in the spectral radius $k.$ Under
   these modest assumptions, therefore, $D_\ell(k)$
   is a normalized density in $k$ and the LES flux $\Pi_\ell$ is a
   ``scale-average'' of the spectral flux $\Pi(k).$ The
   nature of the averaging can be better seen by considering a typical example
   $|\widehat{G}(k)|^2$:

   \vspace{2.5in}

   \noindent A nice example, as pictured, will fall smoothly and monotonically
   from its value 1 near $k=0$ to its
   value 0 at $k=+\infty$ in an interval of width $\approx 1/\ell$ centered at
   $2\pi/\ell.$ From this it follows
   that $D_\ell$ will appear as:

   \vspace{2.5in}

   \noindent Therefore, for a reasonable choice of filter, the function $D_\ell$
   will give an average over
   the interval of width $\approx 1/\ell$ centered at $2\pi/\ell.$

   To study the nature of the cancellations involved with such an average, it is
   convenient to adopt a
   simple model filter $G$ representative of this class. We take
   \be |\widehat{G}(k)|^2=\left\{ \begin{array}{ll}
				1                  & \mbox{ $0\leq k\leq 2\pi$}
   \cr
				2-{{k}\over{2\pi}} & \mbox{ $2\pi\leq k\leq 4\pi$}
   \cr
				0                  & \mbox{ $4\pi\leq k< \infty$}
				\end{array}
			\right. ,\lb{91} \ee
   which is pictured as:

   \vspace{2.5in}

   \noindent In this instance the averaging function $D_\ell$ is just the uniform
   distribution over
   the interval $[2\pi/\ell,4\pi/\ell],$ so that
   \be \Pi_\ell= {{\ell}\over{2\pi}}\cdot
   \int_{2\pi/\ell}^{4\pi/\ell}dk\,\,\Pi(k). \lb{92} \ee
   It was this particular example which we studied in detail in \cite{18} and it
   is essential to the
   arguments of this paper that the reader be familiar with that work.

   We showed in \cite{18} that the
   conventional measure of spectral energy flux, $\Pi(k),$ is inadequate to detect
   the {\em distance}
   energy moves in wavenumber space past the cutoff wavenumber $k.$ By wavenumber
   conservation of the triadic interactions,
   energy may move to at most a wavenumber of magnitude $2k.$ Efficient energy
   transfer of this type is achieved by the
   local triadic interactions, in which two wavenumbers of magnitude near $k$
   transfer energy to a mode with
   wavenumber near $2k.$ By contrast, the Example 1 of \cite{18} achieved a
   transfer to wavenumbers $>k$ by
   simply moving the energy a fixed, small distance $k_0$ outside the sphere of
   radius $k,$ for arbitrarily large
   $k.$ The defect of the traditional flux $\Pi(k)$ is that it does not
   distinguish between these two transfers,
   whereas a proper measure of flux should take into account the {\em relative
   distance} the
   energy is moved outside the sphere. Independent of any connection with LES, it
   was these ideas that led us in
   \cite{18} to introduce the ``band-averaged flux''
   \be
   \overline{\Pi}_\Lambda={{1}\over{2^\Lambda}}
		 \int_{2^\Lambda}^{2^{\Lambda+1}}dk\,\,\Pi(k), \lb{F3} \ee
   as a proper measure of energy transport. This is equivalent to the flux in
   Eq.(\ref{92}) if the filter length
   is chosen there as $\ell=2^{-\Lambda}(2\pi),$ corresponding to a wavenumber
   $2^\Lambda$ at the bottom of the
   $\Lambda$th octave band.

   By a direct argument, it was shown for this quantity in \cite{18} that the
   ``local estimate'' holds
   $ \overline{\Pi}_\Lambda=O\left(2^{\Lambda(1-3h)}\right), $
   corresponding to transfer by local triads. The main property of the
   Navier-Stokes dynamics
   which was used is so-called {\em detailed energy conservation}. ``Partial
   fluxes'' may be
   defined, measuring transfer from the $N$th band to the $L$th, induced by the
   $M$th, as
   \be \Pi^{(M)}_{N\rightarrow L}=\int_\Omega
   d\br\,\,\bv_N(\br)[(\bv_M(\br)\bdot\grad)\bv_L(\br)], \lb{F5} \ee
   in which $\bv_N(\br)$ is the component of the full velocity field obtained by
   summing only over wavenumber
   modes in the $N$th octave band. Simple integration by parts shows that the
   partial fluxes satisfy
   \be \Pi^{(M)}_{N\rightarrow L}+\Pi^{(M)}_{L\rightarrow N}=0, \lb{F6} \ee
   which is the ``detailed conservation'' law. Because of this property, one may
   anticipate that many cancellations
   will occur in the expression Eq.(\ref{F3}) between the small transfers backward
   and forward between neighboring small
   subbands of the $\Lambda$th octave band. The heuristic picture may be explained
   by the
   following figure showing the detailed transfer of energy between subbands of
   size $2^M$ for the contribution
   $\Pi^{(M-1)}(k)$ to the flux induced by the $M$th octave

   \newpage

   \noindent band:

   \vspace{2.5in}

   \noindent This picture suggests that the ``local transfers by nonlocal triads''
   noted by Domaradzki and Rogallo in
   \cite{7} will cancel {\em instantaneously} in the quantity
   $\overline{\Pi}^{(M-1)}_\Lambda.$ The analytical verification
   of these cancellations was given in detail in Section 3 of \cite{18}. The idea
   that such contributions should cancel
   with an appropriate combination of triadic contributions was also pointed out
   independently by Waleffe \cite{11} and
   Zhou \cite{16}. What is important here is that the LES method leads {\em
   automatically} to an appropriate weighted
   average of the flux $\Pi(k)$ for which such cancellations occur.

   Our previous analysis for the model filter Eq.(\ref{91}) actually gives
   somewhat more than just the ``local estimate'' for
   $\Pi_\ell=\overline{\Pi}_\Lambda.$ In fact, our estimates allow us to put
   precise bounds on the fraction of the flux due
   to distant triads. The crucial equations are Eqs.(34) \& (44) of \cite{18}
   which represent the two contributions to the
   flux as sums over the ``deviation parameters'' $\Delta,\Delta',\Delta''$ which
   measure the distance in shellnumber of the
   contributing shells $N,M,L$ from $\Lambda.$ In fact, it is easy to see from
   those equations that the fractional contribution
   to $\overline{\Pi}_\Lambda$ from triads with smallest wavenumber in a shell
   lower than the $(\Lambda-S)$th is
   \be
   {{\overline{\Pi}_\Lambda(S,<)}\over{\overline{\Pi}_\Lambda}}
				  =O\left(2^{-(1-h)S}\right), \lb{108} \ee
   while the contribution from triads with largest wavenumber in higher than the
   $(\Lambda+S)$th shell is
   \be {{\overline{\Pi}_\Lambda(S,>)}\over{\overline{\Pi}_\Lambda}}
			     =O\left(2^{-2hS}\right). \lb{109} \ee
   For the latter estimate, note that the contributions arise from terms with both
   $L>\Lambda$ and $M>\Lambda,$ for which
   $\Delta'$ and $\Delta''$ can differ at most by one because of the wavenumber
   selection rule. This gives the factor of $2$
   in the exponent since $\Delta'\approx\Delta''\approx S.$ The parameter $S$
   measures the degree of nonlocalness of the
   triad (in IR or UV direction), i.e. the minimum number of ``cascade steps'' $S$
   between the smallest and largest wavenumbers.

   Hence, we see that distant triads in fact make an exponentially small
   contribution in shellnumber to instantaneous flux
   when the velocity field has H\"{o}lder type regularity with $0<h<1.$ This is a
   fact well-known for mean-flux as
   calculated within various analytical closures when the energy spectral exponent
   $n$ lies in the range $1<n<3$: see \cite{42a}
   and Appendix 1 \cite{33}. (Note the H\"{o}lder and spectral exponents are
   related heuristically by $n=1+2h.$) However, our results
   here were derived for instantaneous flux in an individual flow realization
   without any closure approximation or statistical assumption.
   The exponent $(1-h)$ gives the ``infrared locality'' whereas the exponent $2h$
   gives the ``ultraviolet locality.''
   The bound $h<1$ seems rather secure but $h>0$ might be violated due to the
   appearance of ``singular structures.''
   It should be appreciated, however, that while the contribution of distant
   triads is exponentially small {\em in shellnumber}
   this corresponds only to a slow, algebraic decay in wavenumber.

   Our accounting of the fractional contribution of distant triads is very close
   to that made by Zhou
   in \cite{9,16}. Our parameter $S$ is essentially the same as his ``scale
   disparity'' $s$ (more properly,
   it corresponds to $\log_2 s$) and
   $\overline{\Pi}_\Lambda(S,<),\overline{\Pi}_\Lambda(S,>)$ are the same as his
   $\Pi^s(k,s),\Pi^e(k,s),$ respectively. In his work, however, he studied {\em
   mean} flux using direct numerical simulation at modest
   Reynolds number or LES to generate the raw statistics. If we take in our bounds
   Eqs.(\ref{108}),(\ref{109})
   above $h={{1}\over{3}}$ (the K41 mean value), then we obtain in both cases
   \be \overline{\Pi}_\Lambda(S)/\overline{\Pi}_\Lambda=O\left(2^{-2S/3}\right),
   \lb{110} \ee
   which Zhou referred to as ``Obukhov scaling'' in the disparity. On the other
   hand, he referred to an estimate
   $O\left(2^{-4S/3}\right)$ as ``Heisenberg scaling.'' In his study, both
   scalings were observed in certain
   cases, although, in our opinion, the inertial-ranges were too short to
   convincingly demonstrate either
   exponent. In any case, it might not be appropriate to directly compare our
   estimates with his since he dealt
   with an average transfer, while our flux estimates are for instantaneous
   transfer in an individual realization.
   Additional cancellations of distant triads could occur in time- or
   ensemble-averaging, so that better
   bounds on the nonlocal contributions could exist. What is important in our
   result is that, given the required
   H\"{o}lder regularity, the ``locality'' of the energy transfer holds at each
   instant for a fixed turbulent flow field.

   \noindent {\em 3.2 Comparison With the EDQNM Study of Leslie-Quarini}

   Superficially similar distinctions as ours above between  transfer for sharp
   Fourier vs. graded filters were observed
   already for mean statistics within analytic closure by Leslie and Quarini
   \cite{LQ}. In fact, it was observed in their
   work that without a low-wavenumber cutoff divergences appear in the sharp
   Fourier case for both forward transfer
   ---``drain''---and backward transfer---``backscatter''--- as the
   grid-wavenumber $K_1$ is approached, verifying an
   earlier observation of Kraichnan based upon the TFM closure \cite{Kr76}. These
   divergences are due to convection by
   vanishingly low wavenumbers, and indicate a breakdown of locality for each of
   the two separate terms. However, the
   divergent parts cancel identically in the net transfer, leaving just a finite
   ``cusp.'' \footnote{In fact, they are
   the same divergences as those observed in Kraichnan's earlier study of energy
   transfer in the ALHDIA closure cited before
   \cite{42}.} It is this cancellation of the divergences due to distant triads
   which justifies our remark above that
   transfer across $K_1$ in the closures is dominated in magnitude by the local
   triads, and {\em not} by the distant triads
   which dominate in individual realizations for the sharp Fourier filter.
   Furthermore, for the Gaussian filter Leslie and
   Quarini found that neither forward nor backward scatter has separately the
   divergence (and even the finite ``cusp''
   essentially disappears.) This last result is apparently in agreement with our
   conclusions, but a closer examination shows
   that there is no real correspondence.

   To make this point we must repeat a little of our previous analysis in the
   context of the closure models.
   We shall follow the notations of Leslie and Quarini \cite{LQ}. Within the
   analytical closures, the mean energy transfer
   into wavenumber $k$ due to interaction of wavenumbers $p,r$ is expressed as
   \be S(k|p,r)=S^{(r)}_{p\rightarrow k}+S^{(p)}_{r\rightarrow k} \lb{131a} \ee
   with
   \be S^{(r)}_{p\rightarrow k}=16\pi^2\cdot\theta_{kpr}\cdot
   k^2p^2r(xy+z^3)q(r)[q(p)-q(k)]. \lb{131b} \ee
   Here, $q(k)=E(k)/4\pi k^2$ is the energy spectrum per Fourier mode, $x,y,z$ are
   geometric
   factors associated to each triad (direction cosines) of order one, and
   $\theta_{kpr}=[\eta(k)+
   \eta(p)+\eta(r)]^{-1}$ is an interaction time-scale associated to the triad.
   Different closure
   models are distinguished by their choice of $\eta(k),$ but a typical one is the
   EDQNM choice
   of $\eta(k)=\nu k^2+\lambda\left(k^3E(k)\right)^{1/2}.$ The average flux is
   expressed in the
   closures by
   \be \Pi(k)=\int_{k}^\infty dk'\,\,\int\int_{\Delta_{k'}}dp\,dr\,\,S(k'|p,r).
   \lb{131d} \ee
   For the locality analysis, the dangerous contribution to consider is
   \be \Pi^{(r)}(k)=\int_{k}^\infty dk'\int_{\Delta_{k'r}}dp
   \,\,S^{(r)}_{p\rightarrow k'} \lb{131e} \ee
   for $r\ll k,$ which arises from convection by the low wavenumber $r.$ It is
   very easy to
   make a simple estimate of this term as
   \begin{eqnarray}
   \Pi^{(r)}(k) & \sim &  \int_{k}^\infty
   dk'\int_{\Delta_{k'r}}dp\,\,\theta_{k'pr}k^2p^2rq(r)[q(p)-q(k')], \cr
	     \, & \sim & r^2\cdot
   k^4r\theta_{rkk}q(r)\cdot\left({{r}\over{k}}\right)q(k). \lb{131f}
   \end{eqnarray}
   The factor of $r^2$ arose from the fact that $k',p$ can only take values in a
   range $r$ around $k.$
   The crucial factor is the $(r/k)$ which arises from expanding $q(k\pm r)$ with
   a power-law spectrum
   and {\em exploiting the cancellation between input and output terms} in
   Eq.(\ref{131b}).

   The energy scaling corresponding to our H\"{o}lder exponent $h$ is $E(k)\sim
   k^{-(1+2h)}.$
   This leads also to $\theta_{rkk}\sim r^{h-1}$ since the time scale for $h<1$ is
   clearly dominated
   by the lowest wavenumber. Hence,
   \be \Pi^{(r)}(k)\sim k^{-2h}\cdot r^{-h}. \lb{131g} \ee
   Since the integral $\int^k_{k_0}dr\,r^{-h}$ is dominated by the UV region when
   $h<1$ and is then
   $\sim k^{1-h},$ it follows finally that
   \be \Pi^{(<)}(k)\equiv \int^k_{k_0}dr\,\Pi^{(r)}(k)\sim k^{1-3h}, \lb{131h} \ee
   exactly as expected. However, if the cancelling in-out contributions were not
   observed
   and the factor of $(r/k)$ missed, then the scaling in Eq.(\ref{131g}) would be
   replaced
   by $\Pi^{(r)}(k)\sim k^{1-2h}\cdot r^{-(1+h)}.$ Since the integral of
   $r^{-(1+h)}$ is
   dominated by the IR region for $h>0,$ the false result $\Pi^{(<)}\sim k^{1-2h}$
   would then
   be obtained. Notice here that, as is well-known, the proper locality estimates
   are obtained
   {\em without any such scale-averaging as we performed previously.} It would be
   possible to
   make such an additional scale average in the closure context, and the same
   cancellations will
   occur since the only requirement for our argument was detailed conservation,
   which is
   satisfied by the ``partial transfers'':
   \be S^{(r)}_{p\rightarrow k}+S^{(r)}_{k\rightarrow p}=0. \lb{131c} \ee
   However, it is clear that the origin of these cancellations will be exactly the
   same as the
   ``in-out cancellations'' without averaging, since in the closures the detailed
   conservation
   condition Eq.(\ref{131c}) is just another consequence of the in-out cancelling
   terms. The
   scale-average is superfluous for the closure flux, which is already local.
   Intuitively, this
   is clear, because the closure flux is supposed to correspond to an
   ensemble-averaged value,
   which will show appropriate cancellations between individual realizations with
   positive and
   negative values of flux.

   We can now comment on the Leslie-Quarini results on graded filters \cite{LQ}.
   In their case, generalizing upon
   Kraichnan in \cite{Kr76}, they defined a wavenumber-dependent eddy viscosity
   $\nu_n(k)=\nu_n(k|K_1)$
   (corresponding to Kraichnan's $\nu(k|k_m)$) as
   \be \nu_n(k)= -\overline{S}(k)/2k^2\overline{E}(k), \lb{131i} \ee
   where they took $\overline{E}(k)=G^2(k)E(k)$ to be the energy spectrum of the
   filtered velocity and
   \be \overline{S}(k)={{1}\over{2}}\int\int_{\Delta_k}dpdr
			  \,\,G^2(k)\{1-G(p)G(r)\}S(k|p,r) \lb{131j} \ee
   was what they defined as the ``true subgrid transfer.'' Again following
   Kraichnan, they decomposed
   the transfer into two terms the ``drain'' part,
   \be 2k^2\nu_d(k)\overline{E}(k)=\sum^{\Delta}{{r}\over{2kp}}
	      \theta_{kpr}b_{kpr}E(r)E(k)\cdot G^2(k)[1-G(p)G(r)] \lb{AA1} \ee
   and the``backscatter'' part
   \be U(k)=\sum^{\Delta}{{k^3}\over{pr}}\theta_{kpr}b_{kpr}E(p)E(r)\cdot
   G^2(k)[1-G(p)G(r)]. \lb{AA2} \ee
   We emphasize that {\em these two terms arise exactly from the output and input
   terms of} $S(k|p,r),$
   respectively. In particular, they do not satisfy separately a detailed
   conservation property such as Eq.(\ref{131c}),
   which arises precisely from a cancellation between those terms. Leslie and
   Quarini then made a corresponding
   division of the eddy viscosity $\nu_n(k)$ into $\nu_d(k),$ the ``drain'' part,
   and $\nu_b(k),$ the
   ``backscatter'' part. They numerically reproduced the result of Kraichnan that
   these two separate
   terms for the Fourier filter each diverges as $k$ approaches the grid
   wavenumber $K_1,$
   and that the divergence cancels in the net viscosity $\nu_n(k).$ However, they
   furthermore
   showed from their numerical work that with the Gaussian filter even the
   separate terms,
   $\nu_d(k)$ and $\nu_b(k)$ are finite for $k$ approaching $K_1.$

   It is clear that this is what one would expect from our analysis. In fact, what
   we have established is much stronger
   than this, because (assuming H\"{o}lder regularity) we showed that the local
   estimate holds for {\em every
   individual realization} with a ``good'' filter, whether instantaneous transfer
   is forward or backward. Since
   the divergences observed by Kraichnan in mean transfer for the sharp Fourier
   case are a consequence
   of locality breakdown, it ought to be anticipated that they will not occur for
   a ``good'' choice of filter.

   However, the kind of cancellations which we established for
   $\overline{\Pi}_\ell$
   in the solutions of Navier-Stokes {\em cannot} occur separately in the
   quantities $\nu_d$ and $\nu_b$
   defined by Leslie-Quarini. In fact, there is no possibility to exploit detailed
   conservation of energy,
   as in our microscopic argument, because in the closure that depends precisely
   upon the
   cancellation between those terms! {\em Therefore, there is no equivalence of
   our results and Leslie-Quarini's.}
   Formally, the disappearance of divergences in $\nu_d,\nu_b$ observed in
   \cite{LQ} appear to arise
   in a quite different way, due to their being simply ``smoothed out'' by the
   averaging functions $G$ in Eq.(\ref{131j}).
   Note that Leslie-Quarini did not give a theoretical explanation of their
   effect---as we have for the
   corresponding phenomenon in Navier-Stokes---but only observed it numerically.
   It seems to be, in fact, a different
   effect. The fact that our argument based upon detailed conservation does not
   work in the closures just
   illustrates the point that the closures do not necessarily reflect reality. In
   fact, it is well known that
   energy conservation does not hold realization by realization in the Langevin
   models of the closures, so that
   the cancellations we established due to detailed conservation in Navier-Stokes
   solutions will not occur there.

   In our opinion part of the difficulty of making a comparison of our result and
   that of Leslie-Quarini lies in the
   different definitions of ``backscatter'' in the two works. In ours here, we
   define a ``backscatter'' event---
   as in current LES practice---as a spacetime point $(\br,t)$ where
   $\Pi_\ell(\br,t)<0.$
   However, the Leslie-Quarini definition of ``backscatter,'' while employing the
   filter function $G,$
   is still essentially Fourier-based. In other words, they define a ``backscatter
   event'' as one where,
   roughly speaking, the mean {\em Fourier} flux $\overline{\Pi}(k)<0$ and then
   take a suitable weighting with the filter
   function. (Even this involves an assumption that the ``input'' term to transfer
   arises from averaging conditioned
   on $\Pi(k)<0$ while the ``output'' term arises from averaging conditioned on
   $\Pi(k)>0$: this is not obviously
   true.) One may expect that these two definitions are globally equivalent, for
   space-averages, but even this does not
   appear easy to demonstrate.

   \section{Conclusions}

   \noindent {\em 4.1 On the ``Multifractal Model''}

   We believe that the present work has demonstrated that the Parisi-Frisch
   ``multifractal model''
   is not just an esoteric theoretical speculation relevant only to exotic
   ``intermittency''
   phenomena in turbulence. In fact, it gets considerable support from the fluid
   equations themselves when
   coupled with some basic facts of turbulence phenomenology. Furthermore, the
   ``multifractal
   model'' provides a coherent framework which rationalizes and explains many
   observations of
   experiment and simulation. Local H\"{o}lder continuity is the {\em maximal}
   regularity consistent
   with the phenomenon of constant mean energy flux through the inertial interval.
   It is also a sufficient
   condition for local energy transfer, necessary for the validity of the
   Kolmogorov
   universality hypothesis. While it must be regarded as tentative and so far
   unproved, we
   believe the ``multifractal model'' is valuable at least as a working
   hypothesis. Further
   investigation, especially mathematical and theoretical, should clarify its
   status.

   \noindent {\em 4.2 On Locality of Transfer}

   The condition that the H\"{o}lder exponent lie in the range $0<h<1$ is
   sufficient to guarantee locality
   of the instantaneous energy transfer. Nonlocal triads with wavevectors in
   distant bands from each other
   contribute an exponentially small fraction of the total flux. However, this
   implies only a
   slow, algebraic decay in Fourier space, so that very high Reynolds numbers
   might be necessary for
   a locally-dominated inertial-range to occur (see also 5.4 below). Furthermore,
   while local transfer
   is certainly necessary for statistical independence of the small scales, it is
   not sufficient. A good
   counterexample is Burgers equation, for which all of our analysis applies and
   in which instantaneous
   energy transfer is likewise local. Nevertheless, the small scales of Burgers
   turbulence do not have
   statistics independent of the large scales but, rather, largely determined by
   them. The key difference
   between Burgers and Navier-Stokes equation is that the former lacks the
   pressure force of the latter
   which tends to stochastically accelerate Lagrangian fluid particles. On the
   contrary, for Burgers
   equation the velocity of a fluid particle is unchanged until it is captured by
   a shock. The difference
   may be characterized as a lack of {\em dynamic locality} for Burgers equation,
   since Lagrangian time
   correlations will not be intrinsic and locally-determined in scale as they are
   for Navier-Stokes, but will
   rather be associated to the overall time of evolution \cite{51a}. These
   considerations show that some
   chaotic features of the Navier-Stokes dynamics---still insufficiently
   well-understood---will be necessary
   to produce independence of successive ``cascade steps''  and universality of
   the small scales.

   \noindent {\em 4.3 On the Filtering Approach}

   The filtering scheme used in the LES modelling technique turns out to be a
   uniquely satisfactory
   method of scale decomposition for turbulent flows, at least in the case where a
   ``good'' filter
   is employed. (For an overview of the ``filtering approach'' to turbulence, see
   the interesting
   paper of Germano \cite{51b}.) The requirements for a ``good'' filter are
   concrete and objective,
   as well as very easily satisfied. With such a filtering technique the basic
   physics of turbulent
   energy transport is made transparent. In contrast, the method of Fourier
   series---despite
   its traditional status in the subject---introduces many subtle and opaque
   effects which
   are just artefacts of using a basis of plane-waves ill-suited to describe the
   physical
   processes involved. A set of interactions like vortex-stretching which are
   naturally
   understood in physical space become difficult to comprehend in a wavenumber
   representation
   and are masked by other effects.

   These same criticisms apply to some apparently more sophisticated
   representations,
   such as {\em wavelet bases} (e.g. see \cite{52}.) Since such bases were
   designed
   to provide a simultaneous space-scale resolution, they might appear ideal for
   turbulence
   applications. However, we have found that the wavelet methods that have so far
   been proposed have the same problems as the Fourier basis and in an even more
   severe
   form! We hope to discuss the wavelet method in detail in a future work
   \cite{53}.

   \noindent {\bf Acknowledgements.} I am very grateful to U. Frisch, R. H.
   Kraichnan, U. Piomelli, and Z.-S. She
   for conversations on various problems discussed in this work, as well as the
   LES group at CTR and two anonymous
   referees for some constructive remarks. I also wish to thank Weinan E for
   pointing out the key identity Eq.(14).
   This research was funded through the NSF Grant No. DMR-93-14938 of Y. Oono and
   N. Goldenfeld, and the support
   of the Physics Department of the University of Illinois at Urbana-Champaign,
   where this work was accomplished.

   \newpage

   \noindent {\bf Appendix: Multifractal Model for Navier-Stokes}

   In Section 2 we reached the conclusion that H\"{o}lder exponents $h\leq 1/3$
   are {\em required} to allow
   for constant mean energy flux in the inertial interval. However, it may seem to
   some that we are engaged in a
   mathematical ``hair-splitting.'' After all---it might be argued---in any
   real flow there is a finite viscosity, no matter how small, and the actual
   velocity field, the solution of viscous
   Navier-Stokes equations, will (presumably) be smooth. However, we would like to
   emphasize that our considerations apply
   even for finite $\nu_0>0$ {\em over the long inertial-range of length-scales},
   when $\nu_0$ is small. It should be appreciated
   that the estimates we have made refer only to the range of scales $\sim \ell$
   and they are unchanged if the velocity-field
   is smooth over much smaller length-scales. To see this clearly, consider the
   difference over length-scales
   $\sim \ell$ of a velocity-field with small-scale components $<a$ removed by
   filtering, when $a\ll \ell.$ If $\bv\in C^\eta,$
   $\eta>0,$ then the filtered field $\vl_a$ is $C^\infty,$ or even analytic, for
   a nice choice of the filter function, and
   $\Delta_{\bl}\vl_a$ will decay rapidly in $\ell$ for $\ell\ll a.$ However,
   \be \Delta_{\bl}\vl_a(\br)\approx\Delta_{\bl}\bv(\br), \lb{37} \ee
   for $\ell\gg a.$ This may be demonstrated analytically using
   \be \Delta_{\bl}\vl_a(\br)=\int
   d\bs\,\,G_a(\bs)\left[\Delta_{\bl+\bs}\bv(\br)-\Delta_{\bs}\bv(\br)\right].
   \lb{38} \ee
   Since only $s\leq a$ contribute to the integral in a good approximation
   (exactly so, if $G_a$ has compact support in the ball $B_a(\br)$ of radius $a$
   centered at $\br$) and $a\ll \ell,$
   $\Delta_{\bl+\bs}\bv(\br)\approx\Delta_{\bl}\bv(\br)$
   and comes out of the integral, whereas the second term is small. A quantitative
   estimate shows that
   \be \Delta_{\bl}\vl_a(\br)=\Delta_{\bl}\bv(\br)+O(a^\eta), \lb{39} \ee
   where $\eta$ may be taken to be the minimum H\"{o}lder exponent of $\bv$ over
   the ball $B_a(\br).$ Therefore, the remainder
   is negligible if $\Delta_{\bl}\bv(\br)\sim \ell^h, a\ll \ell,$ and $h$ and
   $\eta$ are comparable. These arguments
   just support the intuitive idea that $\Delta_{\bl}\bv$ will not be much changed
   if the velocity field is smoothed
   over much smaller scales than $\ell,$ whether by a mathematical operation of
   filtering or by physical regularization due to
   viscosity. In a similar situation, physical Brownian paths are smooth, because
   the size of atoms $a$ is finite, but over
   length-scales $\ell\gg a$ they ``look'' like nowhere-differentiable curves of
   H\"{o}lder index ${{1}\over{2}},$ which
   only becomes exactly true in the idealized limit $a\rightarrow 0.$

   Our previous arguments on energy flux may be given in a form which applies to
   the real situation where Reynolds
   number is finite, but large. Consider a case where turbulence is produced in a
   Navier-Stokes fluid with
   viscosity $\nu_0$ in a box of size $L$ by random forcing at the length-scales
   of the box-size. Suppose, in fact,
   that in every flow realization of the turbulent ensemble
   \be |\Delta_\bl\bv(\br)|\leq v_0\left({{\ell}\over{L}}\right)^h,  \lb{40} \ee
   with some $h>{{1}\over{3}}+\delta$ at each point $\br.$ The constant $v_0$ has
   an interpretation as the typical
   variation of the velocity over the largest eddies of size $\sim L.$ Our
   previous estimates then imply that
   \be |\Pi_\ell(\br)|\leq {{v_0^3}\over{L}}\left({{\ell}\over{L}}\right)^{3h-1},
   \lb{41} \ee
   at each point $\br$ with probability one. However, in that case,
   \be |\langle \Pi_\ell\rangle|\leq {{\langle v_0^3\rangle
   }\over{L}}\left({{\ell}\over{L}}\right)^{3\delta}. \lb{42} \ee
   Let us take $\ell_\nu$ to be a scale intermediate between $L$ and $\lambda,$
   e.g. $\ell_\nu\equiv\sqrt{\lambda L}.$
   Then $\ell_\nu/L\sim ({\rm Re})^{-1/4}$ in the limit $\nu_0\rightarrow 0.$
   However,
   according to our previous discussion, $\langle \Pi_{\ell_\nu}\rangle$ tends to
   a constant $\overline{\varepsilon}$ in the limit
   $\nu_0\rightarrow 0.$ This is not consistent with the previous estimate unless
   $\langle v_0^3\rangle\sim ({\rm Re})^{3\delta/4}.$
   We emphasize that {\em nothing like this is observed.} (At least not in three
   dimensions;  in 2D simulations
   the level of energy fluctuations is indeed observed to rise as
   $\nu_0\rightarrow 0.$) Therefore, the initial
   assumption {\em Eq.(\ref{40}) cannot hold with $\delta>0$ for the range of
   scales} $\lambda\ll\ell\ll L.$

   The ``multifractal model'' accounts for the constant mean flux over this range
   of scales, with $\langle v_0^3\rangle$
   remaining finite, by postulating a suitable distribution of H\"{o}lder
   exponents $h$ in an interval $[h_{\min},h_{\max}],$ so that
   \be \inf_{h\in[h_{\min},h_{\max}]}[(3h-1)+(3-D(h))]=0. \lb{43} \ee
   This is consistent with the requirement from Kolmogorov's
   ``${{4}\over{5}}$-law'' \cite{1c}, which implies
   $\zeta_3=1.$ The exponent $h_*$ for which the infimum is achieved is believed
   to be $\approx {{1}\over{3}}.$
   As stated before, the velocity field could possibly be less regular than this,
   but not more so.

   These arguments imply that actual singularities will occur with some finite
   probability in the turbulent
   ensemble in the limit $\nu_0\rightarrow 0,$ if the currently observed
   finiteness of velocity fluctuations persists
   asymptotically to infinite Reynolds number. However, the present arguments do
   {\em not} require that singularities should
   apppear in finite time for Euler dynamics starting from smooth initial data. In
   fact, we consider a stationary
   turbulent ensemble in which the singularities may have been developed possibly
   over an infinite period of time.
   Even the consideration of decaying turbulence in 3D does not require true
   singularities in finite time. Suppose in
   that case a quasi-equilibrium state is achieved with a long inertial range of
   constant mean flux over the range
   of length-scales $L(t)\gg\ell\gg a(t),$ where $L(t)$ is the (time-dependent)
   integral scale and $a(t)$ is
   a time-dependent lower cutoff to this range. If $a(t)$ were to decrease at a
   faster rate than the rate of decay of energy in the
   large length-scales $\sim L(t),$ say, $a(t)\sim e^{-\gamma t},$ then there
   would be little observable difference from the scenario
   where $a(t)\rightarrow 0$ at a finite time $t=T_*.$ Our previous considerations
   would still imply
   for this case that ``quasi-singularities'' must occur in the velocity field
   over the range of lengths $L(t)\gg \ell\gg  a(t),$
   to be consistent with the constant mean flux in that interval.

   A good example of this is the {\em enstrophy cascade range} for Navier-Stokes
   turbulence in 2D. Considerations of the
   same type as those above show that the enstrophy flux $Z_\ell(\br)$ in 2D obeys
   an estimate
   \be Z_\ell(\br)=O\left(\ell^{2h}\right) \lb{44} \ee
   in terms of the local H\"{o}lder index of the {\em vorticity field}
   $\omega={\bf \nabla\times v}$ at point $\br.$
   In the 2D case the natural ``multifractal'' picture is in fact in terms of the
   H\"{o}lder spectrum of the vorticity field.
   Already in 1967 Kraichnan had proposed that the ultraviolet range of 2D
   turbulence will be an ``enstrophy cascade''
   with a constant mean flux,
   \be \langle Z_\ell\rangle=\overline{\eta}, \lb{45} \ee
   for $\eta_{{\rm Kr}}\ll\ell\ll L$ (here $L=$ the length-scale at which
   enstrophy is injected, $\eta_{{\rm Kr}}=$
   the Kraichnan dissipation length) \cite{32}. From the previous estimate we see
   that
   $h\leq 0$ will therefore be required. This argument is carried out in a
   mathematically rigorous way in
   a separate paper \cite{ME}, especially the last two parts of Section 4.  The
   corresponding argument for the velocity
   regularity $h\leq 1/3$ of Navier-Stokes solutions in the 3D energy
   inertial-range is entirely the same, so that we may
   refer the reader to \cite{ME} for details.

   \end{document}